\documentstyle[aps,twocolumn,epsf]{revtex}

\newcommand{\be}{\begin{eqnarray}}
\newcommand{\ee}{\end{eqnarray}}

\newcommand{\no}{\nonumber}

\renewcommand{\vec}{\bbox}

\begin{document}

\title{Phase diagram of 3D  $SU(3)$ gauge-adjoint Higgs system and C- violation in hot QCD}

\author{S. Bronoff and C. P. Korthals Altes}

\address{Centre Physique Th\'eorique au CNRS, Case 907, Luminy,
F13288, Marseille, France}

\date{\today}

\maketitle

\begin{abstract}
Thermally reduced QCD leads to three dimensional $SU(3)$ gaugefields
coupled to an adjoint scalar field $A_0$. We compute the effective potential
in the one-loop approximation and evaluate the VEV's of $TrA_0^2$ and $TrA_0^3$.
In the Higgs phase not only the former, but also the latter has a VEV.
This happens where the $SU(3)$ gauge symmetry is broken minimally with U(2) still unbroken. The VEV of the cubic invariant breaks charge conjugation and CP. It is plausible that in the Higgs phase one has a transition for large enough Higgs selfcoupling to a region where $TrA_0^3$ has no VEV and where the gaugesymmetry is broken maximally to $U(1)\times U(1)$. For a number of colours larger than 3 an even richer phase structure is possible.
\end{abstract}


\narrowtext


\section{Introduction}
\label{sec1}


To obtain reliable information about the quark-gluon plasma well above the
critical temperature thermally reduced QCD is extremely efficient.
  
Since its inception\cite{dimred} it has been 
pioneered in recent years~\cite{debye} for a precise evaluation 
of the Debye mass for $SU(2)$ and $SU(3)$ gauge groups with or without quarks.

In this note  we want to point out some properties of 
the phasediagram
of $SU(3)$. These properties have to do with the Higgs phase of the three
dimensional gauge-adjoint scalar theory. They are qualitatively different from 
its $SU(2)$ counterpart. 

The main point of this paper is  how and what breaking patterns of the 
various symmetries in the Lagrangian do emerge. R-symmetry breaking ($A_0\to -A_0$) and 
gauge symmetry
breaking are narrowly related, and we surmise where they are realized in 
the phase diagram.  
The Higgs phase in our reduced action is induced by quantum effects.
These effects are calculable for a certain range of values of the parameters 
in the Lagrangian by a loop expansion~\cite{loop}. This calculation
results in a critical line in the phase diagram. Below this line 
the Higgs phase realizes, with the VEV of $TrA_0^2$ non-zero. But in 
$SU(3)$ the VEV of $TrA_0^3$  is not necessarily zero, and, as we will see, 
it acquires a VEV
in the loop expansion. At the same time  the gauge symmetry is minimally
broken, leaving a $U(2)$ group unbroken. This phase is obviously absent in 
$SU(2)$.
However a phase where simultaneously R-parity is restored and the gauge 
symmetry is maximally broken to $U(1)\times U(1)$ is possible and we 
surmise it is realized
in between the symmetric phase and the broken R-parity phase discussed above. 
This would be a phase in which Abelian monopoles screen the two photons
just as in the $SU(2)$ case~\cite{polyakov}. Some aspects of the R-parity 
breaking have been briefly mentioned at LAT98~\cite{altesbron}, and in ref.~\cite{buffa}.

\section{Reduced QCD Lagrangian} 
\label{sec2}


The Lagrangian of reduced QCD reads:
\be
S=\int d\vec x&\{&{1\over 2}\sum_{i,j}TrF_{ij}^2+\sum_iTr(D_iA_0)^2+
  m^2TrA_0^2\no\\
&+&\lambda_1(TrA_0^2)^2+\lambda_2 TrA_0^4+\delta S\}
\label{redaction}
\ee

If the number of colours is 2 or 3, the quartic terms coalesce to one term,
which we will write as $\lambda (Tr{A_0}^2)^2$, with $\lambda=
\lambda_1+\lambda_2 /2$.

The term $\delta S$ stands for all those interactions consistent with gauge,
rotational and R-symmetry (i.e. $A_0\to -A_0$). These are  symmetries
respected by the reduction.

We will study the reduced  action limited  to the superrenormalizable 
terms in eq.~\ref{redaction}, and mostly for $N=3$.
The reduced action has a particular raison d'\^etre: its form is given by 
integrating out the heavy degrees of freedom of QCD at high temperature 
$T$, with coupling $g$ and number of flavours $N_f$. In terms of these 
parameters we have at one loop accuracy~\cite{loop}:
\be
g_3^2=g^2T\no\\
m^2={1\over 3}\left(N+{N_f\over 2}\right)g^2T^2\no\\
\lambda_1={g^4T\over{4\pi^2}}\no\\
\lambda_2=(N-N_f){g^4T\over{12\pi^2}}
\label{para}
\ee

 The gauge coupling $g_3^2$ has dimension of
mass, so the phase diagram is specified by three dimensionless 
parameters~\cite{loop} $x_1=\lambda_1/g_3^2$ ($x_2=\lambda_2/g_3^2$) 
and $y=m^2/g_3^4$. For N=2,3 there are only two such variables, 
and we define $x=\lambda/g_3^2$.
The 4D theory lies on the line determined parametrically by the 
4D coupling g in eq.~\ref{para}, i.e. for $N\ge 4$ and to order $O(x_i)$: 
\be
x_2={1\over 3}(N-N_f)x_1\nonumber\\
yx_1={1\over{12\pi^2}}(N+{N_f\over 2})
\label{parabe4}
\ee
whereas for $N=2,3$:
\be
yx={1\over{72\pi^2}}(N+{N_f\over 2})(6+N-N_f)+O(x)
\label{4D}
\ee

It is obvious that $x\sim g^2$and $xy\sim \mathrm{constant}$ at 
very high temperature. Hence the choice of axes in the phase diagram, fig. 3.

In the reduction process, there is a remarkable cancellation, 
including two loop order, for the coefficients of all renormalizable 
and nonrenormalizable terms containing only $A_0$~\cite{these}.

\section{The effective action}
\label{sec3}

Understanding the phase diagram of the 3D theory necessitates the 
knowledge of the effective action. We define it for SU(3) with its 
quadratic and cubic invariants as:
\be
\label{effaction}
\exp{-VS_{eff}(C,E)}=\int 
DA&\delta&\left(g_3^2C-\overline{TrA_0^2}\right)\no\\
&\delta&\left(g_3^3E-\overline{TrA_0^3}\right)\exp{-S}
\ee

The bar means the  average over the volume V. Hence the volume factor
in front of the effective action on the l.h.s. in eq.~\ref{effaction}.
In the large volume limit we can deduce the effective action for $TrA_0^2$
by minimizing over E, keeping C fixed:

\be
\label{efffaction1}
min_{\{E\}}S_{eff}(C,E)=V(C)
\ee

The effective action for $TrA_0^3$ follows similarly:
\be
\label{efffaction2}
min_{\{C\}}S_{eff}(C,E)=W(E)
\ee

Note we do not use the 1PI functional as is done frequently~\cite{loop}.
Instead we preferred the the potentials $V(C)$ and $W(E)$, since
they are the distributions for the order parameters of relevance
to the phase diagram, hence directly measurable (for earlier
use of manifestly gauge invariant quantities see~\cite{fodor,altes}). 
Taking moments of these distributions gives us the average of $TrA_0^p$, 
$p=2,3$. 

It is very useful to express the effective action $S_{eff}(C,E)$ is  
in terms of the variables $B_1, B_2$ and $B_3$, the diagonal elements 
of a traceless diagonal
$3\times 3$ matrix $B$. This matrix $B$ will be  the background field of 
$A_0$(see section 4), and so, apart from the
fluctuations discussed in the next section:
\be
g_3^2C=TrB^2;\hspace{4mm} g_3^3E=TrB^3
\label{back}
\ee
The matrix $B$ can be expressed in the two diagonal generators of SU(3),
normalised at $1/2$, and the azimuthal angle $\theta$ in the $B$ plane:
{\small\be
B&=&\sqrt 2 C^{1/2}\left(\cos\theta\lambda_8+\sin\theta\lambda_3\right)
\ee}
The two invariants do fix, up to symmetries (see fig. 1), the azimuthal angle:
\be
\cos3\theta=-\sqrt 6 E/C^{3/2}
\label{theta}
\ee

In fig. 1 we have plotted the symmetries of the effective action in this
 plane,
due to permutation symmetry of the $B_i$ and R-symmetry $B\to -B$. As a 
consequence the plane is divided  in sextants. The circle corresponds to
constant $C$.
Also shown are the curves of constant $\vert E\vert$ in 
every sextant, 
and in particular the
asymptotes (dashed lines), where $E=0$. These are the directions where the 
gauge symmetry breaking is maximal, e.g. 
$\lambda_3\sim\mathrm{diag}(1/2,-1/2,0)$. So in the Higgs phase  
with $TrA_0^2\neq 0$ the statement that $TrA_0^3$ has no VEV means 
the symmetry is broken maximally. The opposite is true too: in the 
Higgs phase with maximal gauge symmetry breaking the VEV of the 
cubic invariant is necessarily zero.

\begin{figure}\epsfxsize=7cm
\centerline{\epsfbox{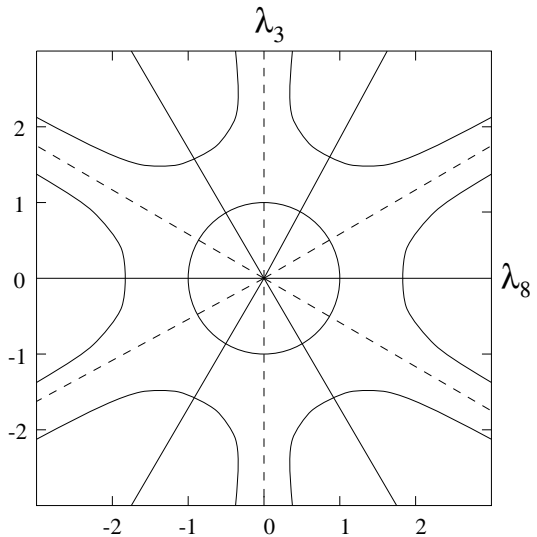}}
\vskip4mm
\caption{ Permutation and R-symmetry render the potential identical
in the six segments between either broken or continuous lines.
The former are directions of maximal, the latter of minimal gauge 
symmetry breaking. The circle is defined by fixing $C$, the cubic 
invariant $\vert E\vert$ fixes the curves with the broken lines as asymptotes.}
\label{fig1}
\end{figure}

Is there an analogous statement for the phase where the cubic invariant 
has a non-zero VEV? That is, does minimal gauge symmetry breaking imply
a VEV for $TrA_0^3$ and vice versa? Look in  fig.1 at the positive 
$\lambda_8$ direction. It is clear that there curves of constant 
$TrA_0^3\neq 0$ and of constant $TrA_0^2,$ have a common tangent 
(in the $\lambda_3$ direction). So if one invariant has a local 
minimum there, so has the other. In section V we show explicitely, 
that both invariants have an absolute minimum in the $\lambda_8$ 
direction. So the answer will shown to be in the affirmative:

In the Higgs phase the cubic invariant is non-zero if and only if 
the gauge symmetry breaking is minimal, i.e. $U(2)$ is still unbroken. 

\section{The Higgs phase through the loop expansion}
\label{sec4}


The loop expansion of $S_{eff}$, eq. 2, will be only trustworthy 
where we find a Higgs phase, hence masses for our propagators. 
In the symmetric phase Linde's argument~\cite{linde} will apply. 
In addition $x$ should be small.
Because the effective potential we use may be less familiar to the reader,
we will go into some detail for the case of SU(3).
The loop expansion starts from the background field B and admits
for small fluctuations around $B$:
\be
A_0&=&B+Q_0\no\\
A_i&=&Q_i
\ee

The gauge invariant constraints have to be taken into account.
Most convenient is to introduce two variables $\gamma$ and $\epsilon$
to Fourier analyse the deltafunction constraints in eq.~\ref{effaction}.
As for the fields, we split them into background and quantum variables:
$\gamma=\gamma_c+\gamma_{qu}$ and likewise for $\epsilon$.
One does do a saddle point approximation around $B$, 
$\gamma_c$ and $\epsilon_c$
 in the path integral in  eq.~\ref{effaction}.
The linear terms in $\gamma_{qu}$ and $\epsilon_{qu}$ give 
eq.~\ref{back}. Linear terms in $\overline{TrBQ_0}$ and 
$\overline{TrB^2Q_0}$ give respectively equations of motion ($V$ is the volume):

\be
-2i\gamma_c +V\left(2m^2+4\lambda TrB^2\right)=0\no\\
\epsilon_c=0
\label{motion2}
\ee
 
The quadratic part in the expansion of $Q_0$ contains a term from the 
constraint
:
\be
-i\gamma_c\overline{TrQ_0^2}+3i\epsilon_c\overline {TrBQ_0^2} 
\ee
apart from the quadratic contribution from the reduced action, 
eq.~\ref{redaction}.
 
Substituting from the equations of motion~(\ref{motion2}) into the 
quadratic part eliminates all terms proportional to $Tr Q_0^2$ and 
$TrBQ_0^2$. The saddle point of this observable ignores the presence of the 
Higgs potential, except that
only massterms proportional to $\lambda (TrBQ_0)^2$ stay. 
So all Higgs masses are zero, except for the diagonal component 
$TrBQ_0$~\footnote{This is in contrast
with methods using the 1PI functional. This will be discussed in a 
later section.} . The only way they still can get masses is through 
the gauge fixing. This is essential for the gauge independence of 
$S_{eff}$, as we shall see now.
 
By choosing $R_{\xi}$ gauge we eliminate mixed terms between $Q_0$ 
and $\vec Q$. The $\xi$ dependence enters only in the masses of the 
off-diagonal ghosts, 
longitudinal vectorbosons and Higgs. For a given off-diagonal component 
(ij), $1\le i\le j\le N$, they  all equal 
$\xi g_3^2(B_i-B_j)^2$. 
Thus integration over these modes gives 
cancellation of the gauge dependent masses.
The diagonal Higgs do not get a mass from  $R_{\xi}$ gauge and the result 
to one loop order reads:
\be
S_{eff}(C,E&)&/g_3^6=yC+xC^2\no\\
&-&{1\over{3\pi}}\{{1\over{g_3^3}}\sum_{1 \le i\le j \le N }\vert 
B_i-B_j\vert^3+2(xC^2)^{3/2}\}
\label{effactionresultt}
\ee
The first line is the tree result, the second line the one loop result.
The transverse gluons contribute the first, the diagonal Higgs component
$TrBQ_0$ the second term. 

This is valid for number of colours $N=2,3$. Going beyond $N=3$ adds a new
invariant $\sim TrA_0^4$ to the reduced action, eq.~\ref{redaction}. 
It also increases the number of constraints in eq.~\ref{effaction}. 
Remarkably, the new coupling does {\it{not}} show up in the one-loop 
result as the reader can easily
work out from the ensuing saddle point equations. Hence the above 
result for $S_{eff}$, supplemented with  the extra tree term 
$\lambda_2 TrB^4$ (that is, the quartic invariant) is valid 
for any N. What is different is the relation of the cubic term, contributed
by the transverse gluons, to the quadratic, cubic, quartic etc. 
invariants, to which we turn now, for the case $N=2$ and $N=3$.

For N=2,  ${\vert B_1-B_2\vert}^3=2\sqrt 2 C^{3/2}$. For N=3
we found it convenient to work with the azimuthal angle $\theta$ in 
fig. 1. The potential for SU(3) becomes then::
\be
S_{eff}(C,E)/g_3^6=yC+xC^2-{2\sqrt 2\over{3\pi}}C^{3/2}a(\theta)
\ee
\be
a(\theta)&=&\vert\sin\theta\vert^3+\vert\sin(\theta+{2\pi\over 3})
\vert^3+\vert\sin(\theta+{4\pi\over 3})\vert^3\no\\
&+&{1\over{\sqrt 2}}x^{3/2}
\label{effactionresult}
\ee

\begin{figure}\epsfxsize=6cm
\centerline{\epsfbox{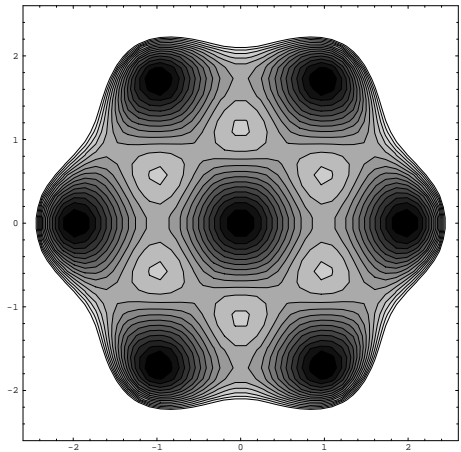}}
\vskip4mm
\caption{The SU(3) potential, eq.(8), on the critical line
as explained in the text, at x=0.1. Horizontal and
vertical axis as in fig. 1.
Dark areas show low, light areas high values. }
\label{fig2}
\end{figure}

In fig. 2 we show a contour plot of the potential for N=3 on the curve 
\be
xy_c={3\over{8\pi^2}}+{\mathcal{O}}(x)
\label{crit}\ee
This is the critical line in the (x,xy) plane of degenerate minima 
along the $\lambda_8$ direction (or  its five equivalents).
The critical line to this order is identical to the 4D physics line for N=3 in eq.~\ref{4D}. Two loop order (or higher) may distinguish them\footnote{ For the two loop result see ~\cite{loop}}.

To obtain the potentials (6) and (7) from our explicit result eq. (15)
we have a simple analytic proof of where the minima are, but  
the reader can see by inspection that both the effective potentials for 
the quadratic and for the cubic invariants are determined by the minimum
along the $\lambda_8$ direction, using eq.~\ref{effaction} and the curves
of constant C and constant E in fig.1.  That is, we minimize the effective
potential ~(\ref{effaction}) with respect to $\theta$ and the 
minima occur at $\theta=0$, mod $2\pi/6$.
As a consequence:

\be
V(C)&=&S_{eff}(C,\theta=0)\no\\
W(E)&=&S_{eff}(C(E),\theta=0)=V(C(E))
\label{endeq}
\ee

In the last equation we used that this direction, $\theta=0$, 
relates the VEV's by:
\be
\vert TrA_0^3\vert={1\over \sqrt 6} (TrA_0^2)^{3/2}
\ee
as follows from eq.~\ref{theta}. Hence where the loop expansion is 
valid the phase with minimal gauge symmetry breaking  and R-symmetry 
breaking realizes.

From the explicit form of the potential for the quadratic VEV, eq.~\ref{endeq}
we find the critical line  
Note 
that on this line
the location of the second minimum is at $C=O(x^{-2})$, and a simple scaling
argument shows, that all terms in the effective action (tree {\it{and}} one loop), 
eq.~\ref{endeq}, are of the {\it{same}} order 
$x^{-3}$. Two loop contributions will be down
by a factor x.

It is helpful to observe that the unbroken U(2) symmetry leaves us
with two off-diagonal zero mass transverse gauge bosons. This is why 
in this direction there is a valley. Up to two loops they give
no problems. They may get a screening mass from the U(2) monopoles in 
this phase, much like the Abelian monopoles screen in the SU(2) case 
the remaining photon~\cite{polyakov}.
So then the stability of this phase would not only be a matter of small
fluctuations around a high VEV (of order $x^{-1}$). There may be an 
instability, when x is so large, that the valley is no longer a 
minimum, and a phase transition to a second Higgs phase will occur.
\begin{figure}\epsfxsize=7.3cm
\centerline{\epsfbox{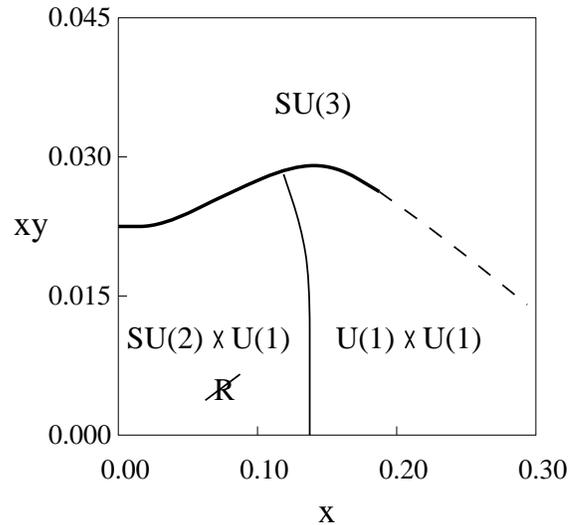}}
\vskip4mm
\caption{Sketch of the SU(3) phase diagram, with unbroken
gaugegroups indicated. R-symmetry is only broken in the left
lower phase}
\label{fig3}
\end{figure}

\section{A second Higgs phase?}
\label{sec5}

Having established the presence of a phase with minimal symmetry
breaking for small x, we have to look for approximation methods 
at large x. One possibility is mean field with a loop expansion
at large values of the lattice coupling and we are analyzing this~\cite{these}.
Also under study is a lattice Montecarlo study~\cite{boucaud}.

On the basis of the final remarks in the preceding paragraph one would 
surmise that transition to come about because the U(2) monopoles become 
unstable and therefore the contribution of the U(2) transverse vectorbosons
grows so large in the $\lambda_8$ direction, that the minimum shifts
to the $\lambda_3$ direction. Along this direction only Abelian monopoles 
survive, and the cubic VEV is zero (see section III).

As we said earlier, this is the phase where Polyakov's mechanism is
at work~\cite{polyakov}. The two photons are screened by the Abelian
fields of the 't Hooft Polyakov monopoles~\cite{thooft}. This phase
would in part of the phase diagram (see fig. 3) join smoothly the symmetric
phase, like in $SU(2)$.

\section{Z(N) invariance and phasestructure for higher N}
\label{sec6}

As noted by the authors of ref.~\cite{loop} the minus sign in front of the 
induced cubic term in the effective action, eq.~\ref{effactionresult}, 
looks formally like  an invariance present in 4D when $N_f=0$ : $Z(N)$ 
invariance.
This invariance involves a large gauge transformation, in the centergroup
 of Z(N), $A_0\sim{2\pi\over N} T/g$.  The reduced theory describes in any of the Z(N) vacua the properties of the plasma at very high T. But it does not
pretend to describe what happens in between two Z(N) vacua, like the tunneling
which leads to the deconfining transition.

Still, formally it is there in eq.~\ref{effactionresultt}, apart from the 
presence of the term $x^{3/2}(TrB^2)^{3/2}$, which for x small can be 
neglected with respect to the
term linear in $x$. Since the invariance is a 4D invariance, we can only 
expect it in the  phasediagram  of the 3D theory, where $x_i$  and $y$
are related to 4D physics, as in eq.~\ref{para} and below.

It is well known, to one loop order, that the Z(N) effective potential
results from eq.~\ref{effactionresultt} by substituting the 4D parameters from
eq.~\ref{para} into it~\cite{loop}:
\be
S_{eff}(y,x)&=& {2\over 3}\pi^2T^3 
\sum_{i,j}\tilde B_{ij}^2(1-\vert\tilde B_{ij}\vert)^2\no\\
&+&O(g^2)
\label{effaction4d}
\ee

The dimensionless field $\tilde B_{ij}$ stands for the difference of 
${g^2\over{2\pi}}(B_i-B_j)/g_3$. It will be of order one and hence the 
dimensionless ratio $B_i/g_3$ is large, of order $g^{-2}\sim x^{-1}$. So the quadratic invariant C will be of order $x^{-2}$, as discussed at  the  end of section IV.

The resulting surface tension $\alpha$~\cite{bhatta} is for 
small $x\sim g^2$ rewritten in 3D language:
\be
{\alpha\over{(g_3^2N)}^2}={(N-1)\over{81\pi^3}}{1\over{\sqrt 2}}
\left({N+6\over{4N}}\right)^{5/2}{1\over{x^{5/2}}}
\label{tension}
\ee
This is valid for N=2 and 3. Note it drops when moving to the right 
on the critical line to larger x, indicating that the transition 
becomes softer. This is as expected from the SU(2) case~\cite{loop}.

It is amusing to go beyond N=3. The phase diagram is then three dimensional 
as discussed below eq.~\ref{para}. Let us take SU(4) as example, with C, E 
and F
 as respectively quadratic, cubic and quartic invariants. The effective 
potential has now a minimum in the direction $\lambda_{15}\sim (1,1,1,-3)$ . The
effective potential for C reads:
\be
V(C)=yC+x_1C^2+x_2 F-{1\over\pi}\left({4\over 3}C\right)^{3/2}
\ee
and the potentials W(E) and X(F) for cubic and quartic invariant have the 
same form as V(C) except for the scaling factors:
\be
E=-1/3\sqrt 3 C^{3/2} ,\hspace{3mm}F=7/12C^2
\ee
In this phase SU(4) breaks to U(3) and the VEV's of all invariants are 
non-zero. There is another possible phase, where the
unbroken subgroup is U(2) and E=0. 
For SU(5) there are 3 possible Higgs phases, one with all four 
invariants a non zero VEV, one with only E=0, and one with only the 
quintic invariant zero.
The critical surface to one loop order is for N arbitrary:
\be
y_c\left(x_1+{(N^2-3N+3)\over {N(N-1)}}x_2\right)
={1\over{36\pi^2}}{N^3\over{(N-1)}}
\ee

This follows immediately from the form of $V(C)$ in the potential valley
in the direction of $\lambda_{N^2-1}$, and solving for the points 
$(x_1,x_2,y)$, where V(C) develops degenerate minima.  
It contains the line of 4D physics in eq.~\ref{parabe4}.

The surface tension now depends on two variables that defines the 
critical surface and  contains as special case the Z(N) surface 
tension on the physical 4D line.

\section{Discussion and comparison with other results}
\label{sec7} 
 
The main finding in this paper, the appearance of a new phase inside
 the Higgs phase with the help of a new order parameter, was leaning
heavily on the use of the distribution function of these order parameters.

Most people practition the 1PI functional, which leads to the same 
results\cite{loop}. as our functional, as we indeed found in this low
order case. It is well known that the difference between the two is
at most in the constant, VEV independent part. As an example of that the 1PI method through
its dependence on the Debye mass reproduces the $O(g^3)$ term of the free
energy in absence of the VEV, whereas our method does not.
The reader may be alarmed that 
in the discussion of the saddle point of our gauge invariant distribution
the y parameter, i.e. the Debye mass, did not appear anymore in the 
propagators of the $A_0$, whereas it does in the propagators
for the 1PI functional\cite{loop}. This feature is just one
of the conditions for it to be gauge independent, as we made amply clear
in $R_{\xi}$ gauge. The only remnant of the Higgs potential in the
propagator of the Higgs is through the quartic coupling, and then only in
the neutral component $TrBQ_0$.

R parity is up to a colour transposition CP in disguise.
The CP transformation in the 4D theory amounts to $A_0\to -A_0^T$ 
(where T stands for matrix transpose), and $A_i\to A_i^T$. Not only the 4D action, but also its reduced version eq.~\ref{redaction}, are invariant under this transformation. So, whenever R-parity is broken, so is CP. In particular,
  on the 4D physics line, eq.~\ref{parabe4} or ~\ref{4D}, we have CP spontaneously broken,
whenever R parity is broken. The same is true for charge conjugation C.

What remains of this phase structure in the original 4D theory? The $A_0$
is the linearized version of the Wilson line. The minimally broken Higgs phase
will then map on the Wilson line $P$ in the $\lambda_8$ directon and connect
to the Z(N) vacua $P=\exp{\pm i{2\pi\over 3}\lambda_8}$. This vacuum will obviously not  break anymore  the gauge symmetry. But C and CP remain broken, because
they map the two Z(3) vacua onto one another. In absence of quarks the $Z(3)$ vacua are mapped onto each other by big gauge transformations.
But this is not true anymore in the presence of quarks, when the Z(N) vacua become non-degenerate, and can form  C or CP violating bubbles. This mechanism and its measurable effects has been discussed in the litterature for strong~\cite{altesk} and electroweak\cite{lee} interactions. It is still subject to discussion because of thermodynamic anomalies~\cite{sem}.

This mechanism might become visible in RHIC physics though C-odd observables involving the energy difference of oppositely charged pions.
Similar observables have been discussed recently~\cite{kharzeev} for a 
mechanism in QCD that creates P-violating bubbles through a quite different
approach.

\section{Summary}
\label{sec8}

In summary, we have shown that at least one Higgs phase is realized with 
spontaneously
broken R-symmetry and minimal gauge symmetry breaking. It is fairly 
plausible that the other phase is located as shown in fig. 3. 
Montecarlo simulations
should test for the additional Higgs phase. 

This phase makes it once more clear, that the Higgs phase does
not correspond to the physical 4D theory, because the broken R-parity
makes it impossible to define a Debye mass! The Debye mass is the 
lowest state with negative R-parity~\cite{arnold} and such a state 
is absent in the new phase.

But in the 4D theory, as argued above, this phase could be a reflection of the C violating properties of the
Z(N) vacua. No such possibility exists for the minimally broken phase, and
the latter is from the point of view of  4D theory an artifact of the reduced theory.

Very interesting is the contrast between the SU(2) and SU(3) monopole
case. What about the non-Abelian monopoles
in the minimally broken phase in SU(3)? Montecarlo studies  as in 
ref.~\cite{teper} should reveal their structure. Some analytic 
progress on classification and dynamics of these monopoles has 
been made by K. Lee and A. Bais~\cite{bais}.
The rich structure for N larger than 3, briefly mentioned at the 
end of section IV, may be relevant for the $SU(5)$ 
transition(see~\cite{rajantie}).

\acknowledgments

The authors thank the ENS, Paris, for its kind hospitality when this work was
done, Raffaele Buffa for help in the early stage of this work and Philippe 
Boucaud for interest and incisive remarks.  C.P.K.A. thanks Sander
Bais, Mike Creutz, Dimitri Kharzeev, Rob Pisarski and Jan Smit for useful discussions.  Mikko Laine and Kari Rummukainen provided us with many
insights. S.B. is indebted
to  the MENESR for financial support.

Note added in proof: K. Kajantie et al.\cite{boucaud} have very recently done
a numerical simulation of the SU(3) phase diagram.

\end{document}